\documentclass{INTERSPEECH2023}

\usepackage{multirow}
\usepackage[flushleft]{threeparttable}
\usepackage{makecell}

\newcommand{\ra}[1]{\renewcommand{\arraystretch}{#1}}

\interspeechcameraready

\title{Handling Trade-Offs in Speech Separation with Sparsely-Gated\\Mixture of Experts}
\name{Xiaofei Wang, Zhuo Chen, Yu Shi, Jian Wu, Naoyuki Kanda, Takuya Yoshioka}
\address{
  Microsoft, One Microsoft Way, Redmond, WA 98052, USA\thanks{The authors thank Robert Gmyr, Felipe Cruz Salinas, Kenichi Kumatani, Yao Qian, Wei Zuo, Jingyan Wang, Devang Patel and Yuan Yu for technical discussions.}}
\email{\{xiaofei.wang, zhuc, yushi, jwu, naoyuki.kanda, tayoshio\}@microsoft.com}

\begin{document}

\maketitle
 
\begin{abstract}
Employing a monaural speech separation (SS) model as a front-end for automatic speech recognition (ASR) involves balancing two kinds of trade-offs. First, while a larger model improves the SS performance, it also requires a higher computational cost. Second, an SS model that is more optimized for handling overlapped speech is likely to introduce more processing artifacts in non-overlapped-speech regions. In this paper, we address these trade-offs with a sparsely-gated mixture-of-experts (MoE) architecture. Comprehensive evaluation results obtained using both simulated and real meeting recordings show that our proposed sparsely-gated MoE SS model achieves superior separation capabilities with less speech distortion, while involving only a marginal run-time cost increase. 

\end{abstract}

\section{Introduction}

Speech separation (SS) offers a solution to handle overlapping voices in automatic speech recognition (ASR)~\cite{yoshioka2018recognizing, watanabe2020chime, raj2021integration}. 
It splits a mixed speech signal into individual clean signals, and thereby allows the ASR system to distinguish the spoken contents of different speakers even when their voices are overlapped.
A great deal of effort has been made to develop effective neural network (NN) architectures for SS~\cite{hershey2016deep, yu2017permutation, wang2018supervised, luo2018tasnet, subakan2021attention, wang2022tf, zhao2023mossformer}.
Recent studies using self-attention-based models, including Transformer~\cite{subakan2021attention} and Conformer~\cite{chen2021continuous}, showed their superior SS capabilities. 
In most studies, the NN models were trained with a limited amount of simulated noisy-clean data pairs, preventing the models from generalizing to unseen domains~\cite{wang2022leveraging}.
Self-supervised learning (SSL) alleviates this problem by using a large amount of unpaired noisy data~\cite{baevski2020wav2vec, hsu2021hubert, chen2022wavlm, huang2022investigating}. Especially, WavLM was shown to significantly improve the SS accuracy~\cite{chang2022end, chen2022continuous}.
With these technical advancements, the SS technology has come to practical use~\cite{yoshioka2019princeton}. 

When we deploy SS models, we face two kinds of trade-offs.
The first trade-off relates to the model capacity and the computational cost.  
While a very large model often improves the SS performance, simply scaling up the model capacity results in a significant increase of the inference cost ~\cite{chen2022continuous}.
The second trade-off pertains to the balance between SS capability and speech distortion.
More specifically, an SS model that is more optimized for overlapping speech often produces distorted signals with more processing artifacts when the input signals contain only one speaker~\cite{chen2021continuous}. 
When SS is used as a front-end for ASR, 
such processing artifacts degrade the ASR accuracy during single-talker regions with high signal-to-noise ratios~\cite{chen2021continuous, sato2021should, iwamoto2022bad, sato2022learning}.

To address these trade-off problems, we explore using a sparsely-gated mixture-of-experts (MoE) architecture~\cite{jacobs1991adaptive, jordan1994hierarchical} for SS. 
MoE with sparse gating provides an increased model capacity with only a marginal run-time cost increase.
To incorporate MoE into the Conformer SS models, we adopt the approach used by Switch Transformer~\cite{fedus2022switch, ma2018modeling}.
Specifically, we replace the feed-forward neural network (FFN) layers of Conformer blocks with multiple parallel FFN layers, or \textit{experts}. 
In each FFN layer, a learnable one-hot gating function determines the expert to be executed for each input during both training and inference.
We conduct comprehensive studies to examine the effectiveness of the MoE SS models and analyze the results from the perspective of the aforementioned two kinds of trade-offs, which were less explored previously. 
Additionally, we analyze the usefulness of a multi-gate MoE (MMoE) architecture. 
To make our evaluation more relevant, we also use embeddings generated by WavLM-pretrained speech models~\cite{chang2022end, chen2022continuous}. 
To the best of our knowledge, this work is the first attempt to incorporate the MoE architectures in the SS models and evaluate their effectiveness thoroughly.

\section{Related work}
\label{sec:relatedwork}
\subsection{Sparsely-gated MoE and MMoE}
\label{sec:prior_moe}

An MoE layer usually consists of $N$ expert networks $E_1$, $\cdots$, $E_N$ and a gating network $G$~\cite{shazeer2017outrageously}. The MoE layer output, $y$, is a weighted combination of $m~(\leqslant N)$ expert outputs, i.e., 
\begin{equation}
    y = \sum_{i \in \mathcal{T}(G(x))} G(x)_i E_i(x),
\end{equation}
where $G$ multiplies input $x$ by a trainable router matrix and calculates the softmax of the product so that 
its $i$th element, $G (x)_i$, represents the gating probability for the $i$th expert $E_i$. 
$\mathcal{T}(G(x))$ returns the indices of the top $m$ experts.
\cite{fedus2022switch} proposed Switch Transformer to further reduce MoE's computational cost by constraining $G$'s output to one-hot activation, where the indices of all the experts are masked except for the one with the largest gating probability. 
The switching MoE mechanism can be used with Transformer and Conformer by replacing each FFN layer of these models with an MoE layer comprising $N$ expert FFN layers of the equivalent size. This significantly increases the model capacity with a small additional run-time cost required for the gating function computation. 

When an MoE model is trained, an auxiliary loss function is often used to avoid an expert under-utilization problem, which occurs when the input data are always routed to a limited number of specific experts. 
An exemplary auxiliary MoE loss as proposed in \cite{fedus2022switch} is defined as follows:
\begin{equation}
\mathcal{L}_\text{MoE} =  \alpha N \sum_{i=1}^{N} f_i P_i,
\label{eq:moe_loss}
\end{equation}
where $f_i$ is the fraction of training samples routed to $E_i$,  
$P_i$ is the fraction of the router probability allocated for the $i$-th expert, and $\alpha$ is the auxiliary loss weight. 
These quantities are calculated within each minibatch. 

MMoE extends MoE to multi-task learning~\cite{ma2018modeling}. 
It uses multiple gating networks optimized for individual tasks while sharing the experts across the tasks. 
In this paper, 
we utilize the MMoE architecture to bias an SS model to not produce processing artifacts.

\subsection{Monaural speech separation (SS)}
\label{sec:ss}
When a mixture of multiple speech signals is observed, SS attempts to isolate the individual signals. 
We adopt a time-frequency (T-F) masking approach.  
That is, the SS model, parameterized by $\theta$ as $\text{SS}({\bf Y}; \theta)$, takes in the short-time Fourier transform (STFT) magnitude features, ${\bf Y} \in \mathbb{R}^{T \times F}$, derived from the observed signal. Then, it outputs T-F masks, ${\bf M}_1$, $\cdots$, ${\bf M}_S$, for all speakers, where ${\bf M}_i \in \mathbb{R}^{T \times F}$. 
Note that $T$ and $F$ denote the numbers of time frames and frequency bins, respectively. 
The $i$-th speech signal is estimated by applying inverse STFT to the element-wise multiplication of ${\bf M}_i$ and ${\bf Y}$. 

The SS model is optimized to minimize an utterance-level permutation invariant training (uPIT) loss. 
When we measure the loss in the mel frequency domain, it is defined as:
\begin{equation}
\mathcal{L}_{\textrm{uPIT}} = \min_{p \in P} \sum_{i = 1}^{S}  \left \Vert \textrm{Mel}({\bf M}_i \otimes|{\bf Y}|) - \textrm{Mel}(|{\bf X}_{j}|)\ \right \Vert  _2^2, 
\label{eq:upit}
\end{equation}
where $(p_1, \cdots, p_S)$ is a permutation of $(1, \cdots, S)$, $P$ is the set of all possible permutations, operator $\otimes$ denotes element-wise multiplication,  $|{\bf X}_j|$ is the magnitude of the $j$-th reference speech, and $\textrm{Mel}(\cdot)$ denotes the mel-filterbank transform. 

To employ the trained SS model as an ASR front-end to pre-process long-form conversations, continuous speech separation (CSS) was introduced~\cite{yoshioka2018recognizing, wu2021investigation, chen2021continuous}.
CSS uses a sliding window (e.g., a 2.4-s window with a 0.8-s hop) and performs SS at each window location. The output signals from each window position are stitched together across the windows~\cite{chen2021continuous} in a way that maintains the speaker consistency between neighboring windows. 
Since utterance overlaps involving three or more speakers are rare in real meetings~\cite{cetin06_interspeech}, the maximum number of simultaneously active speakers, $S$, can be set to 2. 

\begin{figure}[t!]
  \centering
  \includegraphics[width=.5\textwidth]{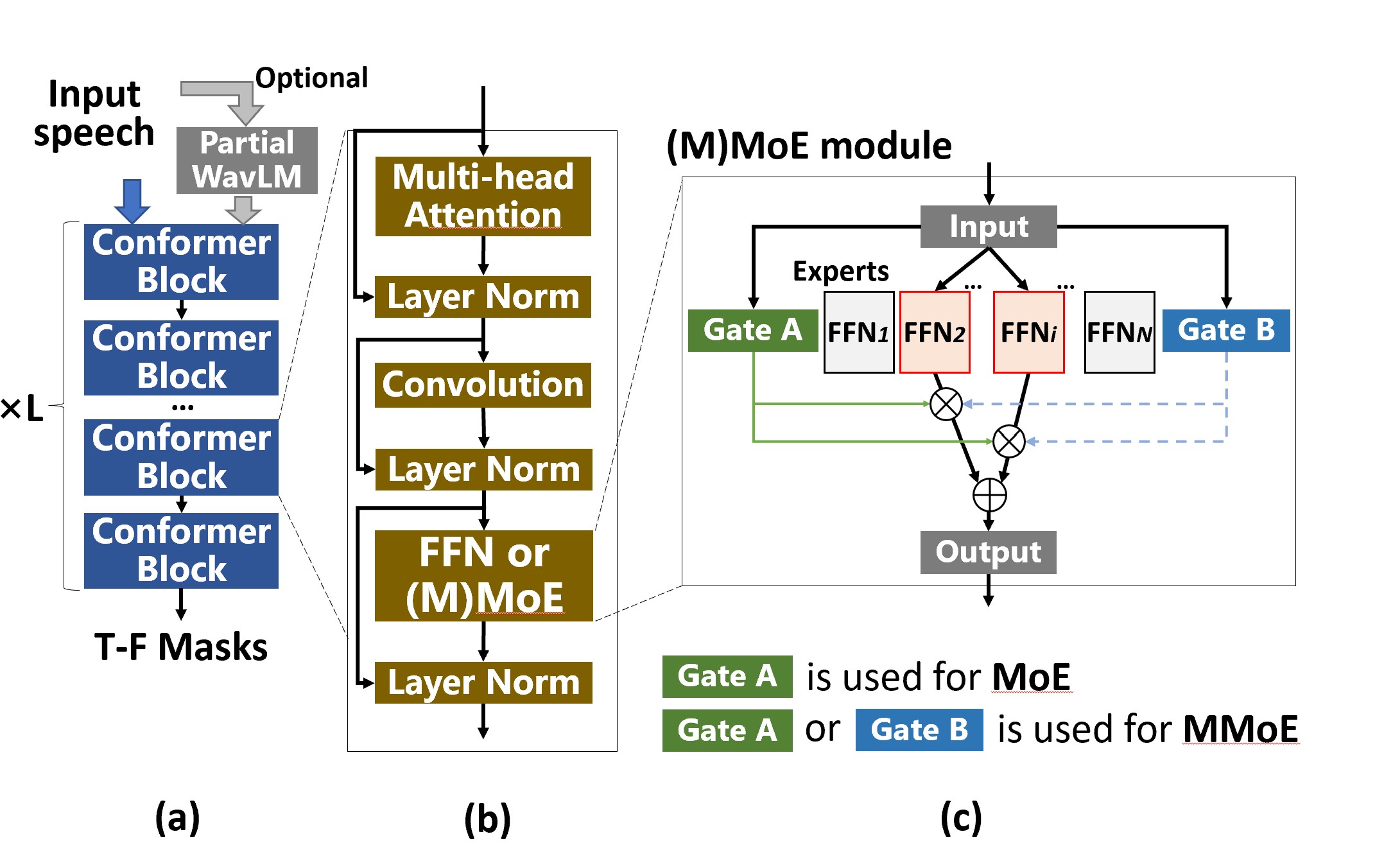}
  \vspace{-3mm}
      \caption{SS model with sparsely gated architecture. (a) Conformer-based SS model with optional WavLM embeddings; (b) our Conformer block; (c) architecture of single-gate (gate A only) or multi-gate (gates A and B) MoE.}
      \label{fig:overview}
\end{figure}

We use an SS model consisting of a stack of Conformer blocks (see Fig.~\ref{fig:overview}(a)) due to its high SS capability as demonstrated in \cite{wu2021investigation}. Furthermore, in order to make our investigation more practically relevant, we additionally utilize a speech representation obtained from a vast amount of unpaired noisy data with unsupervised learning~\cite{hsu2021hubert, chen2022wavlm, huang2022investigating}.
Here, we use a pre-trained WavLM model. 
Following~\cite{chen2022continuous}, for each input sample, we extract layer-wise speech representations of the pre-trained model. Then, we calculate their weighted average with layer-wise learnable weights and append it to the STFT magnitude features to create an input to the Conformer SS model. 
Based on the observation in \cite{chen2022wavlm}, we can include only the first several layers in the weighted average computation, which helps moderate the increase in the computational cost.

\section{SS with sparsely-gated MoE}
\label{sec:method}
Section \ref{sec:moe} describes how we apply MoE to an SS model to increase the model capacity without significantly increasing the inference cost. 
Section \ref{sec:mmoe} extends MoE to MMoE with the aim of reducing processing artifacts when processing non-overlapped speech. 

\subsection{Conformer-based SS with MoE}
\label{sec:moe}
Fig.~\ref{fig:overview} shows an overview of our Conformer-based SS model with MoE.
We use a variant of the Conformer block as shown in Fig.~\ref{fig:overview}(b).
Our Conformer block consists of a multi-head self-attention module with relative position encoding, a convolution module, and a dense FFN module, where each module has a residual connection and layer normalization. No dropout layers are included within the Conformer block. 

To introduce MoE into the Conformer SS model, we replace FFN modules of some selected Conformer blocks with MoE modules as shown in Fig.~\ref{fig:overview}(c). Each expert in the MoE module consists of a linear layer, a ReLU activation layer, a dropout layer, and another linear layer. The shapes, and thus the computational costs, of the linear layers remain the same as those of the FFN module of the non-MoE model. 
In addition to the experts, we use a gating function that multiplies the input by a learnable linear matrix and calculates the softmax of the product to calculate the routing probabilities for the experts. 
To save computational cost, we use top-1 routing for the expert selection during both training and inference.
Starting from the first Conformer block, we apply the MoE module for every other Conformer block.
The model is trained by combining the loss functions of 
Eqs. \eqref{eq:moe_loss} and \eqref{eq:upit}, i.e., 
\begin{equation}
\mathcal{L}_\textrm{SS-MoE} = \mathcal{L}_\textrm{MoE} + \mathcal{L}_{\textrm{uPIT}}.
\label{eq:loss}
\end{equation}

\subsection{Conformer-based SS with MMoE}
\label{sec:mmoe}
We further extend MoE to MMoE to cope with the trade-off between the SS accuracy and the processing artifacts observed during non-overlapping speech regions. 
Prior studies in acoustic noise reduction~\cite{sato2021should,sato2022learning} demonstrated that the negative impact of the processing artifacts on the ASR accuracy could be mitigated by selectively disabling the noise reduction processing. 
Inspired by this finding, we apply a different gating function for non-overlapped speech to promote 
some experts to focus more on non-overlapped speech and thus learn to introduce less speech distortion. 

To realize this idea, we utilize an MMoE with two gating functions that calculate the gating probabilities for the shared experts differently. 
First, we group the training samples into two classes: one consisting of samples including overlapped speech, and the other consisting of samples that do not contain overlapped speech. 
When we create a training minibatch, we choose training samples exclusively from only one class. 
Then, we use the gating function that corresponds to the class of the current mini-batch. 
Specifically, as illustrated in Fig.~\ref{fig:overview}(c), we use Gate A when processing an overlapped-speech minibatch whereas Gate B is employed for a minibatch that does not contain overlapped-speech samples. 
We apply voice activity detection to individual reference (i.e., pre-mix) signals to determine the presence of speech overlaps for each training sample. 

At inference time, we simply use the gating function learned from the non-overlapped-speech minibatches, 
instead of introducing an additional speech overlap detector to switch between the two gating functions. 
We choose to do so in order to avoid incurring an additional computational cost and to bias the SS processing towards not introducing processing artifacts. 

\section{Experimental results}
\label{sec:exp}

As one of our objectives is to reduce SS's negative impact on ASR due to the processing artifacts, 
we conducted ASR experiments to evaluate the proposed MoE-based SS models.

\subsection{SS model configurations}
\label{sec:ss_moe_config}

We used two classes of Conformer SS models. 
Models of the first class made use of 257-dimensional magnitude spectra as input. They had 18 Conformer encoder layers, each comprising an 8-head and 512-dimension self-attention layer, a 1024-dimension FFN layer, and a convolution layer with a kernel size of 33 and a channel number of 512. 
The models had 59M parameters in total. 

Models of the second class utilized WavLM-based embeddings as additional input features by following the configuration of \cite{chen2022wavlm}. 
We used three WavLM models. The first one consisted of 24 layers with 317M parameters,
which we refer to as WavLM X-Large. The second one used the bottom 8 layers of X-Large, resulting in 103M model parameters. We call it WavLM Large. 
The third model, called WavLM Small, consisted of 8 layers with a total of 23M parameters. 
The first two WavLM models produced 1024-dimensional embeddings, resulting in a 1281-dimensional input for the SS models, whereas WavLM Small yielded 384-dimensional embeddings.  
When we used the WavLM models, we employed smaller Conformer SS models to keep the total run-time cost within a reasonable range. Specifically, we used 16 Conformer layers, each with 4 attention heads and 256 dimensions. 
The total parameter count was 26M. 

\subsection{Training configuration}

We trained our SS models on 10k hours of data generated by following the recipe of \cite{yoshioka2018recognizing}. Our training dataset consisted of a balanced combination of single-speaker utterances and two-speaker mixtures, with four kinds of overlap patterns~\cite{wu2021investigation, yoshioka2022vararray}, so that the models can deal with various speech-overlap conditions. 
We used VoxCeleb~\cite{nagrani2020voxceleb}, WSJ, LibriSpeech~\cite{panayotov2015librispeech} and our in-house data to obtain clean speech signals and mixed the signals to create the 10k-hour training dataset. 
We added room impulse responses randomly generated with the image method~\cite{allen1979image} and noise samples obtained from the DNS Challenge~\cite{reddy2021icassp} and the MUSAN noise datasets~\cite{snyder2015musan}. 

All models were trained with 8 NVIDIA A100 GPUs, an AdamW optimizer with a 10-3 weight decay, 
a minibatch size of 512 and a learning rate (lr) scheduler with a linear warm-up and decay. 
Each training sample was 4-second long. 
The peak lr was set to 1e-4, and the total update and warm-up steps were 300k and 30k, respectively.

For the MoE model training, we distributed the experts to different GPUs. 
The capacity factor~\cite{fedus2022switch} and the expert dropout rate were set to 1.5 and 0.1, respectively. The switching jitters, which controls the multiplicative fraction to the gate input, and $\alpha$ were set to 0.01~\cite{microsoft2022ortmoe}.

\subsection{Simulation experiments}
\label{sec:utt_eval}

First, we conducted experiments with an utterance-based SS setting by using artificially mixed test samples to assess the usefulness of the MoE architecture under controlled conditions. 

\subsubsection{Evaluation data and method}

Our simulated evaluation dataset consisted of single-talker (i.e., non-overlapped) utterances and their artificial mixtures. These utterances were sampled from single-talker regions of real meeting recordings that we collected in various meeting rooms. 
They were recorded by either close-talking microphones or distant microphones. 
The quantities of the respective recording types were approximately 100k and 300k words. Accordingly, we ended up with the following four subsets:
\begin{itemize}
\item {\bf Close Clean:} non-overlapped utterances recorded with close-talking microphones; 
\item {\bf Far Clean:} non-overlapped utterances recorded with distant microphones; 
\item {\bf Close Mix:} overlapped utterances obtained by mixing randomly chosen two Close Clean samples; 
\item {\bf Far Mix:} overlapped utterances obtained by mixing randomly chosen two Far Clean samples. 
\end{itemize}
The overlap ratio was set to 40 \% for the latter two subsets. 
Close Mix and Far Mix had the same number of words as Close Clean and Far Clean, respectively. 
We used the latter two subsets to measure the SS performance. The former two subsets were used to assess the impact of processing artifacts.

For each test sample, we applied an SS model to the entire audio to obtain two separated signals. Then, we performed ASR for each signal with an in-house hybrid ASR model [{\it citation}]. 
To calculate the word error rates (WERs) for each test sample, we matched up the reference transcriptions with the ASR outputs in a way that minimized the average WER. 

In addition, we measured the real-time factors (RTFs) of different SS models. 
We ran each model to process a 2.4-second sample 100 times and calculated the average of the elapsed times. 
We used an Intel Xeon W-2133 3.60GHz Processor with a single-thread setup.

\begin{table}[t]
  \centering
  \caption{WERs (\%) of STFT feature-based SS models for artificially-mixed evaluation dataset.}
  \ra{0.9}
  \tabcolsep = 1.5mm
  \label{tab:exp_conformer}
  \vspace{-3mm}
{\footnotesize
  \begin{tabular}{@{}ccccccccc@{}}
  \toprule
   \multirow{3}{*}{SS Model} & \multirow{3}{*}{\makecell{Number\\ of Experts\\ (\#Param.)}} & \multirow{3}{*}{ \makecell{RTF \\ for SS} } && \multicolumn{5}{c}{ WER (\%)} \\ 
  &                   &      && \multicolumn{2}{c}{ Close}   && \multicolumn{2}{c}{Far}   \\ 
    \cmidrule{5-6} \cmidrule{8-9}
  &              &      && { Mix}    & { Clean} && { Mix}  & { Clean} \\
  \midrule
  None    & -  &  -                    && 91.9 & 12.8 && 86.8 & 13.5 \\
  \midrule
   Dense               & - (59M)                       &  0.130 && 23.5 & 20.9 && 23.0 & 15.0 \\ 
     MoE              &  4 (87M)                &  0.132 && 21.9 & {\bf 17.3} && {\bf 21.9} & {\bf 14.9} \\ 
     MoE              &  8 (125M)                     &  0.136 && {\bf 21.4} & 17.7 && 22.7 & 15.0 \\ 
     MoE              &  16 (201M)                    &  0.141 && 21.5 & 17.6 && 22.1 & 15.1 \\ 
  \bottomrule 
\end{tabular}
}
\end{table}

\begin{table}[t]
  \centering
  \caption{WERs (\%) of WavLM-based SS models for artificially mixed evaluation dataset.}
  \ra{0.9}
  \tabcolsep = 0.7mm
  \label{tab:exp_wavlm}
  \vspace{-3mm}
  {\footnotesize
  \begin{tabular}{@{}cccccccccc@{}}
  \toprule
  \multirow{3}{*}{SS Model} & \multirow{3}{*}{\makecell{WavLM\\(\#Param.)}} & \multirow{3}{*}{\makecell{Number\\of Experts\\(\#Param.)}}  & \multirow{3}{*}{RTF} && \multicolumn{5}{c}{WER (\%)} \\ 
   &    &          &      && \multicolumn{2}{c}{Close}   && \multicolumn{2}{c}{Far}    \\ 
   \cmidrule{6-7} \cmidrule{9-10}
   &   &            &      && {Mix}    & {Clean} && {Mix}  & {Clean} \\
  \midrule
   Dense & Large (105M) & - (26M) &  0.310 && 20.3 & 19.5 && 20.9 &  15.1 \\
   MoE   & Large (105M) & 4 (43M)          &  0.313 && 18.1 & 17.5 && 20.5 & 15.1 \\
   MoE   & Large (105M) & 8 (60M)              &  0.316 && {\bf 16.7} & {\bf 16.8} && {\bf 19.6} & {\bf 15.0}  \\
   \midrule
   Dense & Small (23M) & - (26M) &  0.094 && 21.7 & 20.9 && 21.6 & 14.9 \\
   MoE   & Small (23M) & 4 (43M) &  0.095 && {\bf 18.1} & {\bf 17.5} && {\bf 20.5} & 15.1 \\
   MoE   & Small (23M) &  8 (60M) &  0.097 && 19.0 & 17.9 && 21.6 & {\bf 14.8} \\
   \bottomrule 
\end{tabular}
}
\end{table}

\subsubsection{Results and discussions}

Table~\ref{tab:exp_conformer} compares a dense SS model using STFT magnitude features and three MoE models with different expert numbers.
We can see that the dense model significantly improved the WERs for both Close Mix and Far Mix, as expected. However, it also severely degraded the WERs for Close Clean and Far Clean due to the artifacts that were generated by the SS processing. 
The MoE SS models outperformed the dense model for all conditions, except for the 16-expert model increasing the WER for Far Clean by a 0.1 percentage point. 
The performance gains were pronounced for the close microphone recording conditions. 
It is noteworthy that these improvements were obtained with modest RTF increases in the range of 1.5\%-8.5\% relative. 
These results show the advantage of the sparsely-gated MoE models. 
Since the 16-expert model provided no clear benefits in all the conditions, we did not further use this setting in the following experiments. 

Table \ref{tab:exp_wavlm} shows the experimental results obtained by using WavLM Large and Small embeddings. 
We can see the same performance trend as was observed in the experiment using the STFT magnitude features as input. 
Moreover, the MoE SS models using WavLM Small outperformed those using the STFT magnitude features while reducing the RTFs, which demonstrates the effectiveness of combining the MoE architecture and the features obtained with unsupervised learning. 

\begin{figure}[t]
  \centering
  \includegraphics[width=.45\textwidth]{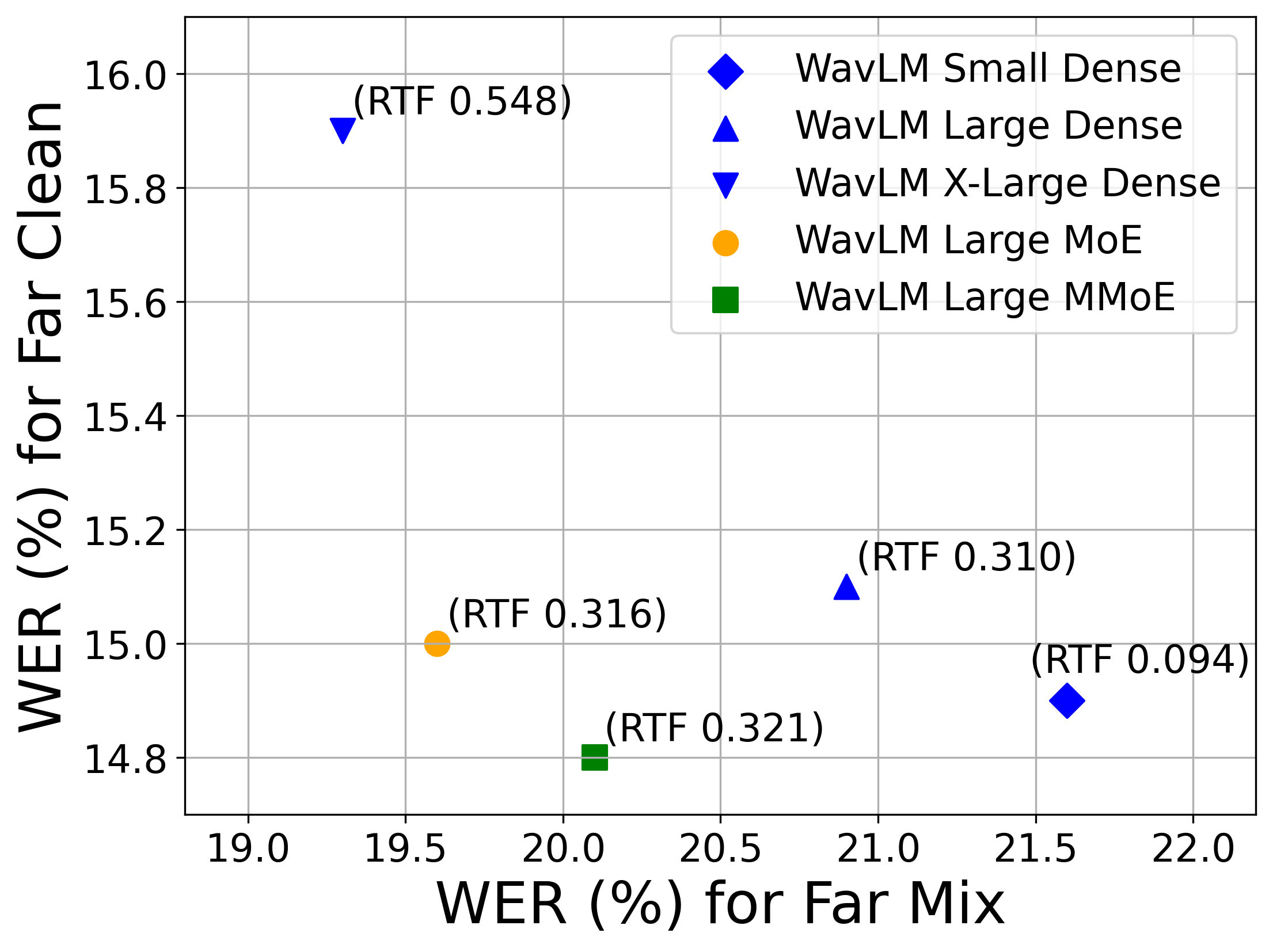}
      \caption{WERs (\%) for ``Far Clean'' and ``Far Mix'' for various WavLM-based SS models. The numbers next to data points represent the RTF of each SS model.}
      \label{fig:exp_mmoe}
\end{figure}

To visualize the trade-off between the SS capability and speech distortion, we plot different WavLM SS models in a 2D coordinate plane with x-axis and y-axis being the WERs for Far Mix and Far Clean, respectively, 
in Fig.~\ref{fig:exp_mmoe}. 
If we compare the three blue data points, which represent dense models, 
we can see that using larger WavLM models improved the SS capability at the expense of the performance for the Far Clean condition, implying the existence of the performance trade-off. 
In contrast, both MoE (orange) and MMoE (green) achieved better trade-off points than the dense models (blue). Note that we used 8-expert models in this plot. 
We can also see that MMoE outperformed MoE in the Far Clean condition, suggesting that it produced less speech distortion. This supports our hypothesis that the multi-gate approach further promotes some experts being optimized for dealing with non-overlapping speech to suppress speech distortion.
However, this performance gain was obtained at the expense of the performance for Far Mix. 
From this analysis, we may conclude that, while using MMoE did not further mitigate the trade-off problem, it provided another way to manage the trade-off by encouraging some experts to optimize themselves for non-overlapped speech. 

\subsection{Real meeting experiment}

To evaluate the effectiveness of the MoE-based SS models in practical settings, 
we further carried out an experiment using real meeting recordings.  
We utilized  the single distant microphone recordings of the AMI~\cite{carletta2005ami} and ICSI~\cite{janin2003icsi} datasets.
We segmented the audio signals by reference silence positions, following \cite{kanda2021large}. 
For each segment, which contains one or more speakers, we applied CSS by using an SS model obtained in the previous section. 
Then, we applied our hybrid ASR system to each separated signal and measured 
the speaker-agnostic WER \cite{fiscus2006multiple}. 
For CSS, we used a sliding window of 2.4 seconds with a hop size of 0.8 seconds. 

Table~\ref{tab:css_eval_rtf} shows the WERs of the dense and 4-expert MoE SS models using the WavLM Small embeddings. 
Using the MoE architecture consistently improved the WER for all test sets, while it required a less than 1\% RTF increase (see Table \ref{tab:exp_wavlm}). 
This result confirms the usefulness of the MoE architecture even for the real meeting data under the CSS scheme. 

\begin{table}[h]
\caption{WER (\%) for AMI and ICSI test sets. WavLM Small embeddings were used as SS model input.}
  \ra{0.9}
  \tabcolsep = 0.5mm
\label{tab:css_eval_rtf}
\centering
{  \footnotesize
\begin{tabular}{@{}ccccc@{}}
\toprule
\multirow{2}{*}{SS Model} & \multicolumn{4}{c}{WER (\%)}  \\ 
\cmidrule{2-5}
& {AMI-dev} & {AMI-eval} & {ICSI-dev} & {ICSI-eval}  \\ 
\midrule
Dense & 19.8 & 23.0 & 18.6 &  17.3    \\ 
MoE (4 experts) & {\bf 19.4} & {\bf 22.4} & {\bf 18.4} & {\bf 17.1}   \\ 
\bottomrule 
\end{tabular}
}
\vspace{-3mm}
\end{table}

\section{Conclusions}
We described SS models based on sparsely-gated MoE architectures and experimentally explored its effectiveness. 
Extensive evaluations were conducted by using both simulated and real datasets. 
Our ASR results and RTF analysis showed that using the sparsely-gated MoE  architecture can efficiently increase the SS model capacity and provide gains in both SS accuracy and speech distortion. Using MMoE was shown to offer an additional way to control the trade-off between the SS accuracy and speech distortion. The best performance-to-cost trade-off was attained when we combined a small WavLM model with the MoE SS model. 
\bibliographystyle{IEEEtran}
\bibliography{mybib}

\end{document}